\documentclass[12pt]{article}

\newcommand{\be}{\begin{equation}}
\newcommand{\ee}{\end{equation}}
\newcommand{\nopass}{\ensuremath{\mathit{nopass}}}
\newcommand{\wep}{\ensuremath{\mathit{WEP}}}

\usepackage{amssymb,amsfonts,amsmath}
\usepackage{times}
\usepackage{natbib}
\usepackage{fancyhdr}
\usepackage{multirow}
\usepackage[pdftex]{graphicx}
\usepackage{epstopdf}
\usepackage{latexsym}
\usepackage{amsbsy}
\usepackage[usenames]{color}
\usepackage{caption}
\begin{document}

\bibliographystyle{nature}

\renewcommand{\multirowsetup}{\centering}

\title{WiFi Epidemiology: Can Your Neighbors' Router Make Yours Sick?}

\author{Hao Hu$^{1,2}$, Steven Myers$^{2,*}$, Vittoria Colizza$^{2,3}$, Alessandro Vespignani$^{2,3}$}
\maketitle

\vspace{-0.6cm}
\noindent {\small{$^1$Department of Physics, Indiana
University, Bloomington, IN 47405, USA}}

\noindent {\small{$^2$School of Informatics, Indiana University,
Bloomington, IN 47406, USA}}

\noindent {\small{$^3$Complex Networks Lagrange Laboratory,
Institute for Scientific Interchange (ISI), Torino 10133, Italy}}

\noindent {\small{$^*$e-mail: samyers@indiana.edu}}
\vspace{0.3cm}\\
\noindent {\bf In densely populated urban areas
WiFi routers form a tightly interconnected proximity network that
can be exploited as a substrate for the spreading of malware able to
launch massive fraudulent attack and affect entire urban areas WiFi
networks. In this paper we consider several scenarios for the
deployment of malware that spreads solely over the wireless channel
of major urban areas in the US. We develop an epidemiological model
that takes into consideration prevalent security flaws on these
routers. The spread of such a contagion is simulated on real-world
data for geo-referenced wireless routers. We uncover a major
weakness of WiFi networks in that most of the simulated scenarios
show tens of thousands of routers infected in as little time as two
weeks, with the majority of the infections occurring in the first 24
to 48 hours. We indicate possible containment and prevention measure
to limit the eventual harm of such an attack.}\\

\noindent Correspondence and requests for material should be
addressed to S.M. (samyers@indiana.edu) or A.V. (alexv@indiana.edu).

\newpage
The use of WiFi routers is becoming close to mainstream in the US
and Europe, with 8.4\% and 7.9\% of \emph{all} such respective
households having deployed such routers by 2006 \cite{wifiusage},
and a WiFi market expected to grow quickly in the next few years as
more new digital home devices are being shipped with WiFi
technology.  In 2006 alone 200 million WiFi chipsets were shipped
worldwide, representing nearly half of the  500 million cumulative
total \cite{wifimarket}. As the WiFi deployment becomes more and
more pervasive, the larger is the risk that massive attacks
exploiting the WiFi security weaknesses could affect large numbers
of users.

Recent years have witnessed a change in the designers behind a
malware attack and in their motivations, corresponding to the ever
increasing sophistication needed to bypass newly developed security
technologies. Malware creators have shifted from programmer
enthusiasts attempting to get peer credit from the ``hacker''
community, to organized crime engaging in fraud and money laundering
through different forms of online crime. In this context WiFi
routers represent valuable targets when compared to the PC's that
malware traditionally infect, as they have several differing
properties representing strong incentives for the attacks. They are
the perfect platform to launch a number of fraudulent
attacks~\cite{driveby-pharm,GO06,MJ07,chrisblog} that previous
security technologies have reasonably assumed were
unlikely~\cite{trawlerPhishing}. Unlike PCs, they tend to be always
on and connected to the Internet, and currently there is no software
aimed at specifically detecting or preventing their infection.
Further, as routers need to be within relatively close proximity to
each other to communicate wirelessly, an attack can now take
advantage of the increasing density of WiFi routers in urban areas
that creates large geographical networks where the malware can
propagate undisturbed. Indeed, many WiFi security threats  have been
downplayed based on the belief that the physical proximity needed
for the potential attack to occur would represent an obstacle for
attackers. The presence nowadays of  large ad-hoc networks of
routers make these vulnerabilities considerably more risky than
previously believed.

Here we assess for the first time the vulnerability of WiFi networks
of different US cities by simulating the wireless propagation of
malware, a malicious worm spreading directly from wireless router to
wireless router. We construct an epidemiological model that takes
into account several widely known and prevalent weaknesses in
commonly deployed WiFi routers' security~\cite{driveby-pharm,
warkitting}, (e.g., default and poor password selection and cracks
in the WEP cryptographic protocol~\cite{nailwep}). The WiFi
proximity networks over which the attack is simulated are obtained
from real-world geographic location data for wireless routers. The
infection scenarios obtained for a variety of US urban areas are
troublesome in that the infection of a small number of routers in
most of these cities can lead to the infection of tens of thousands
of routers in a week, with most of the infection occurring in the
first 24 hours. We address quantitatively the behavior of the
spreading process and we provide specific suggestions to minimize
the WiFi network weakness and mitigate an eventual attack.

\bigskip
\noindent {\bf Results and Discussion}

\bigskip
{\it WiFi networks}. WiFi routers, even if generally deployed
without a global organizing principle, define a self-organized
proximity communication network. Indeed, any two routers which are
in the range of each other's WiFi signal can exchange information
and may define an ad-hoc communication network. These networks
belong to the class of \emph{spatial} or \emph{ geometric networks}
in that nodes are embedded in a metric space and the interaction
between two nodes strongly depends on the range of their spatial
interaction~\cite{Dall:2002,Nemeth:2003,Herrmann:2003,Helmy:2003}.

\begin{figure}[!ht]
\begin{center}
\vskip .7cm
\includegraphics[width=15cm]{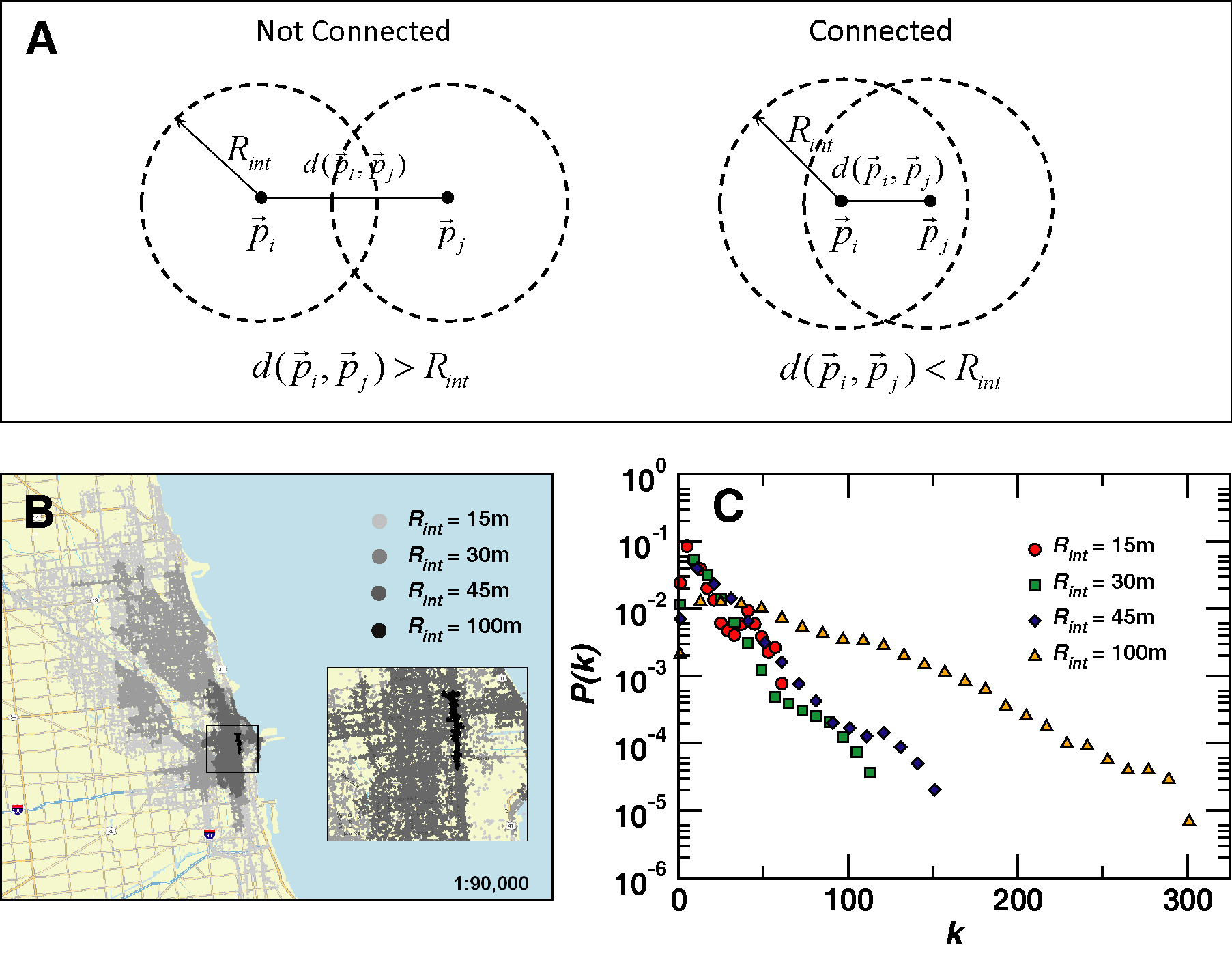}
\vskip .1cm
{\bf Figure 1.}
\caption{\label{fig:GC}}
\end{center}
\end{figure}

In this perspective, one might wonder if the actual deployment of
WiFi routers is sufficient at the moment to generate large connected
networks spanning sizeable geographic areas. This problem,
equivalent to the percolation of giant connected component in graph
theory\cite{Molloy:1995,Bollobas:2006}, is however constrained by
the urban area's topology and demographic distribution
 dictating the  geographical locations of WiFi routers. Here we
consider WiFi networks as obtained from the public worldwide
database of the Wireless Geographic Logging Engine (WiGLE)
website~\cite{wigle}. The database  collects data on the worldwide
geographic location of wireless routers and counts
more than $10$ million unique networks on just under 600 million
observations~\cite{wigle_data}, providing good coverage of the
wireless networks in the United States and in North Central Europe.
The data provide a wealth of information that include, among other
things, the routers' geographic locations (expressed in latitude
$LAT$ and longitude $LON$) and their encryption statuses. In
particular, we focused on the wireless data extracted from seven
urban areas or regions within the United States~--~ Chicago, Boston,
New York City, San Francisco Bay Area, Seattle, and  Northern and
Southern Indiana. Starting from the set of vertices corresponding to
geo-referenced routers in a given region, we construct the proximity
network~\cite{Dall:2002,Nemeth:2003,Herrmann:2003,Helmy:2003}
 by drawing an edge between any two routers $i$ and $j$
located at $\vec{p}_i=(LON_i,LAT_i)$ and $\vec{p}_j=(LON_j,LAT_j)$,
respectively, whose geographical distance $d(\vec{p}_i,\vec{p}_j)$
is smaller than the maximum interaction radius $R_{int}$ (i.e.,
$d(\vec{p}_i,\vec{p}_j) \le R_{int}$), as shown in
Figure~\ref{fig:GC}A. In the WiFi networks, the maximum interaction
radius $R_{int}$ strongly depends on the local environment of any
specific router. In practice, $R_{int}$ ranges from  15m for a
closed office with poor transmission to approximately 100m
outdoors~\cite{wifibook}. For simplicity, we assume that $R_{int}$
is constant, independent of the actual location of a given router,
and we consider four different values of the maximum interaction
radius~---~ $R_{int}\in\{\mbox{15m, 30m, 45m, 100m}\}$~---~analyzing
the resulting networks for each of the seven regions under study. A
more detailed account of the network construction procedure and the
filtering methods used to minimize potential biases introduced by
the data collection mechanisms are described in the Materials and
Method section.

In Figure~\ref{fig:GC}B we report an illustration of the giant
component of the network obtained in the Chicago area for different
values of $R_{int}$. It is possible to observe that despite the
clear geographical embedding and the city constraints, a large
network of more than 48,000 routers spans the downtown area for
$R_{int}$  set to 45 meters. The degree distributions of the giant
components, reported in Figure~\ref{fig:GC}C, are characterized by an
exponential decrease~\cite{Herrmann:2003} with a cutoff clearly
increasing with the interaction radius, since a larger range
increases the number $k$ of nodes found within the signal area. Very
similar properties are observed in all the networks analyzed. It is
important to stress that the metric space embedding exerts a strong
preventative force on the small-world behavior of the WiFi networks,
since the limited WiFi interaction rules out the possibility of long
range connections.

\bigskip

{\it Infecting a Router}. The infection of a susceptible router
occurs when the malware of an already infected router is able to
interface with the susceptible's administrative interface over the
wireless channel. Two main technologies  aim at preventing such
infection through i) the use of encrypted and authenticated wireless
channel communication through the WEP and WPA cryptographic
protocols, and ii) the use of a standard password for access
control. The encryption should provide a higher level of security,
as it needs to be bypassed before a potential attacker could attempt
to enter the router's password. Most users do not currently employ
their routers encryption capabilities~--~indeed the  encryption
rates in the considered cities vary from 21\% to 40\% of the
population. For the purposes of this work we assume that WPA is not
vulnerable to attack\footnote{This is not completely accurate, see
supplementary discussion for a more in depth discussion.}, and
therefore any router that uses it is considered \emph{immune} to the
worm. Because of cryptographic flaws in WEP, this  protocol can
always be broken given that the attacker has access to enough
encrypted communication. This can be achieved by waiting for the
router to be used by legitimate clients, or by deploying more
advanced active attacks. Bypassing WEP encryption is therefore
feasible and only requires a given amount of time.

Once the malware has bypassed any cryptographic protocol and
established a communication channel, it may then attempt to bypass
the password. A large percentage of users do not change their
password from the default established by the router manufacturer,
and these passwords are easily obtainable. For  legal reasons it is
difficult to measure exactly what this percentage is, so here we use
as a proxy the percentage of users who do not change their routers
default SSID. For all the other routers, we assume that $25\%$ of
them can have the password guessed with 65,000 login attempts, based
on the evidence provided by security studies~\cite{klein90foiling}
which showed that $\approx 25\%$ of all users passwords are
contained in a dictionary of 65,000 words. We then pessimistically
assume, based on previous worms, that another $11\%$ of passwords
are contained in a larger library of approximately a million words
\cite{jeff00memorability}.  No backoff mechanism exists on the
routers that prevents systematic dictionary attacks.  In case the
password is not found in either dictionary, the attack cannot
proceed. Alternatively, if the password has been overcome, the
attacker can upload the worm's code into the router's firmware, a
process that typically takes just a few minutes.  In the Material
and Methods section we report a list of the typical time scales
related to each step of the attack strategy.

\bigskip
{\it Construction of the epidemic model}. The construction of the
wireless router network defines the population and the related
connectivity pattern over which the epidemic will spread. In order
to describe the dynamical evolution of the epidemic (i.e., the
number of infected routers in the population as a function of time)
we  use a framework analogous to epidemic modeling that assumes that
each individual (i.e. each router) in the population is in a given
class depending on the stage of the infection~\cite{Anderson:1992}.
Generally, the basic modeling approaches consider three classes of
individuals: susceptible (those who can contract the infection),
infectious (those who contracted the infection and are contagious),
and recovered (those who recovered or are immune from the disease
and cannot be infected). In our case the heterogeneity of the WiFi
router population in terms of security attributes calls for an
enlarged scheme that takes into account the differences in the
users' security settings. We consider three basic levels of security
and identify the corresponding classes: routers with no encryption,
which are obviously the most exposed to the attack, are mapped into
a first type of susceptible class $S$; routers with WEP encryption,
which provides a certain level of protection that can be eventually
overcome with enough time, are mapped into a second type of
susceptible class denoted $S_{\wep}$; routers with WPA encryption,
which are assumed to resist any type of attacks, correspond to the
removed class $R$. This classification however needs to be refined
to take into account the password settings of the users that range
from a default password to weak or strong passwords and finally to
non-crackable passwords. For this reason, we can think of the
non-encrypted class $S$ as being subdivided into four subclasses.
First, we distinguish between the routers with default password
$S_{\nopass}$ and the ones with password $S_{pass1}$. The latter
contains routers with all sorts of passwords that undergo the first
stage of the attack which employs the smaller dictionary. If this
strategy fails, the routers are then classified as $S_{pass2}$ and
undergo the attack which employs the larger dictionary. Finally, if
the password is unbreakable, the router is classified as
$R_{hidden}$. The last class represents routers whose password
cannot be bypassed. However, their immune condition is \emph{hidden}
in that it is known only to the attacker who failed in the attempt,
while for all the others the router appears in the susceptible class
as it was in its original state. This allows us to model the
unsuccessful attack attempts of other routers in the dynamics. WEP
encrypted routers have the same properties in terms of password, but
the password relevance starts only when the WEP encryption has been
broken on the router. At this stage of the attack it can be
considered to be in the non-encrypted state, and therefore no
subclasses of $S_{\wep}$ have to be defined. In addition to the
above classes, the model includes the infected class ($I$) with
those routers which have been infected by the malware and have the
ability to spread it to other routers.

The model dynamics are specified by the transition rates among
different classes for routers under attack. Transitions will occur
only if a router is attacked and can be described as a reaction
process. For instance the infection of a non-encrypted router with
no password is represented by the process $S_{\nopass}+I\to 2I$. The
transition rates are all expressed as the inverse of the average
time needed to complete the attack. In the above case the average
time of the infection process is $\tau=5$ minutes and the
corresponding rate $\beta$ for the transition $S_{\nopass}+I\to 2I$
is $\beta=\tau^{-1}$. Similarly the time scale $\tau_{\wep}$ needed
to break a WEP encryption will define the rate $\beta_{\wep}$ ruling
the transition from the $S_{\wep}$ to the non-encrypted class. In
the Materials and Methods section we report in detail all the
transition processes and the associated rates defining the epidemic
processes.

One of the most common approaches to the study of epidemic processes
is to use deterministic differential equations based on the
assumption that individuals mix homogeneously in the population,
each of them  potentially in contact with every
other~\cite{Anderson:1992}. In our case, the static non-mobile
nature of wireless routers and their geographical embedding make
this assumption completely inadequate, showing the need to study the
epidemic dynamics by explicitly considering the underlying contact
pattern~\cite{Watts:1998,Barabasi:1999,Keeling:1999,Moore:2000,PastorSatorras:2001}.
 For this reason, we rely on numerical simulations obtained
by using an individual-based modeling strategy. At each time step
the stochastic disease dynamics is applied to each router by
considering the actual state of the router and those of its
neighbors as defined by the actual connectivity pattern of the
network. It is then possible to measure the evolution of the number
of infected individuals and keep track of the epidemic progression
at the level of single routers. In addition, given the stochastic
nature of the model, different initial conditions can be used to
obtain different evolution scenarios.

As multiple seed attacks are likely we report simulations with
initial conditions set with $5$ infected routers randomly
distributed within the population under study. Single seed attacks
and different number of initial seeds have similar effects. The
initial state of each router is directly given by the real WiFi data
or is obtained from estimates based on real data, as detailed in the
Material and Methods section. Finally, for each scenario we perform
averages over 100 realizations.

\bigskip

{\it Spreading of synthetic epidemics}.  According to the simulation
procedure outlined above we study the behavior of synthetic
epidemics in the seven urban areas we used to characterize the
properties of WiFi router networks. The urban areas considered are
quite diverse in that they range from a relatively small college
town as West Lafayette (Indiana) to big metropolis such as New York
city and Chicago. In each urban area we focus on the giant component
of the network obtained with a given $R_{int}$ that may vary
consistently in size.

Here we report the results for a typical epidemic spreading
scenario in which the time scales of the processes are chosen
according to their average estimates. The best and worst case
scenarios could also be obtained by considering the combination of
parameters that maximize and minimize the rate of success of each
attack process, respectively. The networks used as substrate  are
obtained in the intermediate interaction range of 45m.

\begin{figure}[!ht]
\begin{center}
\vskip .7cm
\includegraphics[width=15cm]{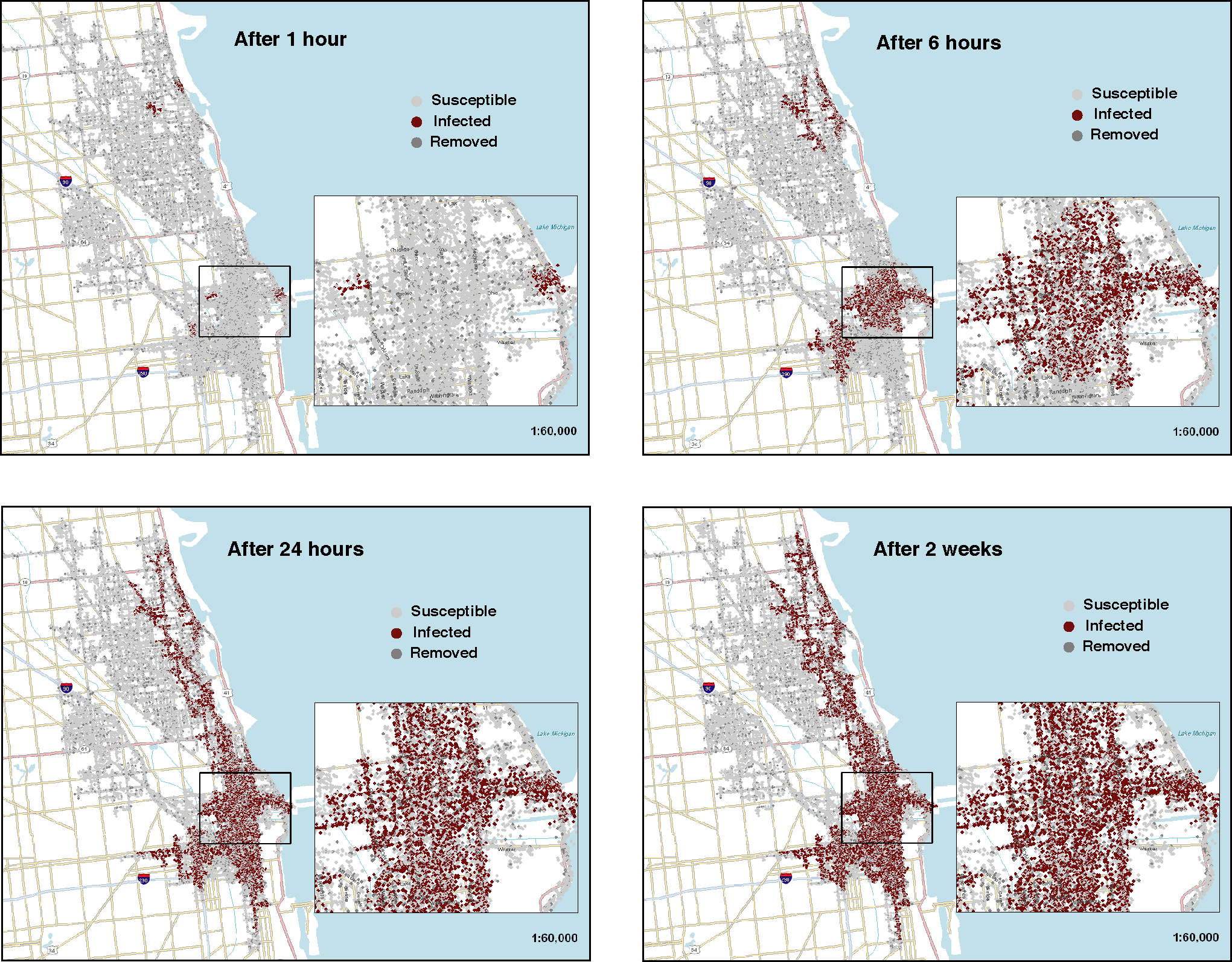}
\vskip .1cm
\caption{\label{fig:ChicagoPanels}}
{\bf Figure 2.}

\end{center}
\end{figure}

\begin{figure}[!ht]
\begin{center}
\vskip .7cm
\includegraphics[width=15cm]{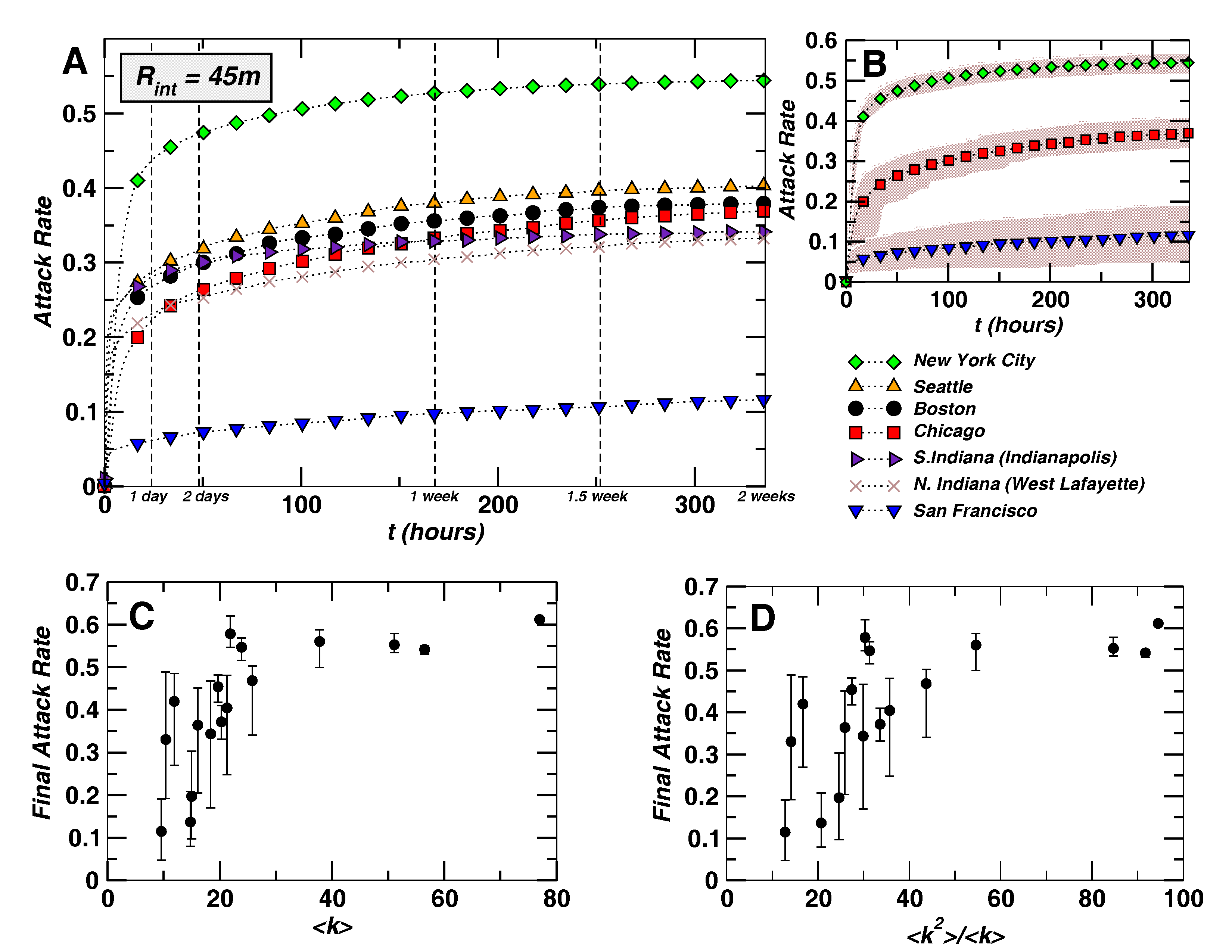}
\vskip .1cm
\caption{\label{fig:ARvsTime}}
{\bf Figure 3.}

\end{center}
\end{figure}

The four snapshots of Figure~\ref{fig:ChicagoPanels} provide an
illustration of the evolution of a synthetic epidemic in the Chicago
area; shown in red are the routers which are progressively infected
by malware. The striking observation is that the malware rapidly
propagates on the WiFi network in the first few hours,  taking
control of about 37\% of the routers after two weeks from the
infection of the first router. The quantitative evidence of the
potential impact of the epidemic is reported in
Figure~\ref{fig:ARvsTime}A-B, where the average profile of the density
of infected routers is reported for all the urban areas considered
in the numerical experiment, together with the corresponding fluctuations.
While it is possible to notice a
considerable difference among the various urban areas, in all cases
we observe a sharp rise of the epidemic within the first couple of
days and then a slower increase, which after two weeks leaves about
10\% to 55\% of the routers in the giant component controlled by
malware. The similar time scale in the rise of the epidemic in
different urban areas is not surprising as it is mainly determined
by the time scale of the specific attacks considered in the malware
spreading model. In general the sharp rise of the epidemic in its
early stages is due to the non-encrypted routers which are infected
in a very short time. The slower progression at later stages is
instead due to the progressive infection of WEP routers whose attack
time scale is about one order of magnitude longer. Single
realization results clearly show the effect of the interplay of
different time scales involved in the spreading phenomenon.

A more complicated issue is understanding the different attack
(infection) rates that the epidemic attains in different urban area
networks. The pervasiveness of the epidemic can be seen as a
percolation effect on the WiFi
network~\cite{Grassberger:1983,Ben-Avraham:2000}. The WPA encrypted
routers and those with unbreakable passwords represent obstacles to
the percolation process and define an effective percolation
probability that has to be compared with the intrinsic percolation
threshold of the
network\cite{Ben-Avraham:2000,Cohen:2000,Callaway:2000}. The larger
the effective percolation probability with respect to the threshold,
the larger the final density of infected routers. On the other hand,
the epidemic thresholds of the networks are not easy to estimate
because they are embedded in the particular geometries of the
cities' geographies. In random networks, large average degree and
large degree fluctuations favor the spreading of epidemics and tend
to reduce the network percolation
threshold~\cite{PastorSatorras:2001,Lloyd:2001}.
Figure~\ref{fig:ARvsTime}C-D shows an appreciable statistical
correlation between the attack rate and these quantities. On the
other hand, there are many other network features that affect the
percolation properties of the networks. First, the cities have
different fractions of encrypted routers. While these fraction are
not extremely dissimilar, it is clear that given the non-linear
effect close to the percolation threshold, small differences may
lead to large difference in the final attack rate. For instance, San
Francisco, with the largest fraction of encrypted routers
corresponding to about 40\% of the population, exhibits the smallest
attack rate amongst all the urban areas considered. Second, the
geometrical constraints imposed by the urban area geography may have
a large impact on the percolation threshold, which can be rather
sensitive to the local graph topology. For instance, network layouts
with one dimensional bottlenecks or locally very sparse connectivity
may consistently lower the attack rate by sealing part of the
network, and thus protecting it from the epidemic. Indeed, a few WPA
routers at key bottlenecks can make entire subnetworks of the giant
component impenetrable to the malware.

The present results offer general quantitative conclusions  on the
impact and threat offered by the WiFi malware spreading in different
areas, whereas the impact of specific geographical properties of
each urban area on the epidemic pattern will be the object of
further studies.

\bigskip
\noindent {\bf Conclusions}

\bigskip
Based on this work, we note that there is a real concern about the
wireless spread of WiFi based malware. This suggests that action
needs to be taken to detect and prevent such outbreaks, as well as
more thoughtful planning for the security of future wireless
devices, so that such scenarios do not occur or worsen with future
technology. For instance, given the increasing popularity of
802.11n, with its increased wireless communications range, the
possibility for larger infections to occur is heightened, due to the
larger connected components that will emerge. Further, many devices
such as printers, and DVR systems now ship with 802.11 radios in
them, and if they are programmable, as many are, they become
vulnerable in manners similar to routers. Lastly, it is highly
likely that we will only see the proliferation of more wireless
standards as time goes by, and all of these standards should
consider the possibility of such epidemics.

There are two preventive actions that can be easily considered to
successfully  reduce the rates of infection. First, force users to
change default passwords, and secondly the adoption of WPA, the
cryptographic protocol meant to replace WEP that does not share its
cryptographic weaknesses. Unfortunately, the dangers of poorly
chosen user passwords have been widely publicized for at least two
decades now, and there has been little evidence of a change in the
public's behavior. In addition, there are many barriers to public
adoption of WPA on wireless routers. The use of only one device in
the home that does not support WPA, but that does support the more
widely implemented WEP, is sufficient to encourage people to use WEP
at home. However, unlike the more traditional realm of internet
malware, the lack of a small world contagion graph implies that
small increases in the deployment of WPA or strong passwords can
significantly reduce the size of a contagion graph's largest
connected component, significantly limiting the impact of such
malware. In future work we plan a detailed study of the percolation
threshold of the giant component as a function of the proportion of
WPA immune nodes in order to provide quantitative estimates for the
systems' immunization thresholds.

\bigskip
\noindent {\bf Materials and Methods}

\bigskip
{\it WiFi Data and Networks}. WiFi data is downloaded from the WiGLE
website~\cite{wigle} for seven urban areas in the US and is
processed in order to eliminate potential biases introduced by data
collection. Records that appear as \emph{probe} in their type
classification are removed from the dataset since they correspond to
wireless signals originating from non-routers. Such records
represent a very small percentage of the total number in every city
considered. For example, in the Chicago urban area there were 4433
\emph{probe} records, corresponding to 3.7\% of the total.

A preliminary spatial analysis of the data for each urban area
reveals the presence of sets of several WiFi routers sharing an
identical geographic location. In order to avoid biases due to
overrepresentation of records, we checked for unique BSSID (i.e.,
MAC address) and assume that each of these locations could contain
at most $n$ overlapping routers, where $n$ was fixed to 20 to
provide a realistic scenario, such as a building with several
hotspots. For the Chicago urban area, this procedure led to the
elimination of 3194 records, which represent 2.7\% of the total
number of WiFi routers.

More importantly, we adopt a randomized procedure to redefine the
position of each router in a circle of radius $R_{ran}$ centered on
the GPS coordinates provided by the original data. This procedure is
applied to approximate the actual location of each router which
would be otherwise localized along city streets, due to an artifact
of the wardriving data collection method~\cite{wigle,wardriving}.
The newly randomized positions for the set of routers completely
determine the connectivity pattern of the spatial WiFi network and
its giant component substrate for the epidemic simulation. Results
presented here are obtained as 5 averages over several randomization
procedures. Table~\ref{table_properties} reports the main
topological indicators of the giant components  of each urban area
extracted from the WiFi network built assuming $R_{int}=45$m.

\begin{figure}[!ht]
\begin{center}
\vskip .7cm
\includegraphics[width=8cm]{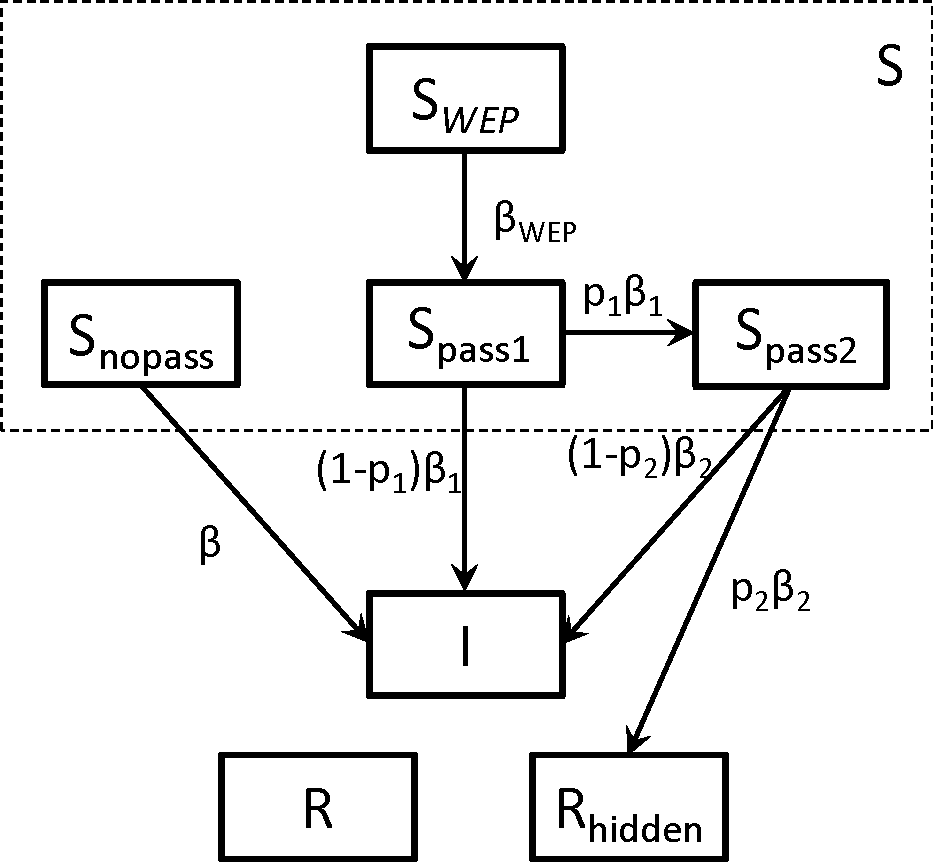}
\vskip .1cm
\caption{\label{fig:compartment_flow}}
{\bf Figure 4.}

\end{center}
\end{figure}

\bigskip
{\it Epidemic model}. Figure~\ref{fig:compartment_flow} shows the
flow diagram of the transmission model.

Initial conditions set the number of routers belonging to each of
the following compartments: $S_{nopass}$ (routers with no encryption
and default password), $S_{pass1}$ (routers with no encryption and
user set password),  $S_{\wep}$ (routers with WEP encryption), and
$R$ (routers with WPA encryption, here considered immune). The
classes $S_{pass2}$ and $R_{hidden}$ are void at the beginning of
the simulations since they represent following stages of the
infection dynamics. Encrypted routers are identified from original
data, and the fraction of $R$ out of the total number of encrypted
routers is assumed to be $30\%$, in agreement with estimates on real
world WPA usage. Analogously, we assume that the non-encrypted
routers are distributed according to the following proportions: 50\%
in class $S_{nopass}$ and 50\% in class $S_{pass1}$.

The infection dynamics proceeds as follows. A router with no
encryption enters the infectious class with unitary rate $\beta$ if
attacked. The attack to a router in class $S_{pass1}$ is
characterized by a transition rate $\beta_1$ and has two possible
outcomes: with probability $(1-p_1)$ the router is infected and
enters $I$, whereas with probability $p_1$ it enters $S_{pass2}$
since the attacker is not able to overcome the password and the
infection attempt requires additional time and resources. Once in
class $S_{pass2}$, it can become infectious with  probability
$(1-p_2)$ if the attack is successful, or otherwise the router
enters $R_{hidden}$ with probability $p_2$ because the password has
not been bypassed. This process occurs with a transition rate
$\beta_2$. WEP encrypted routers follow the same dynamics once the
encryption is broken and they enter  $S_{pass1}$ with transition
rate $\beta_{WEP}$.

The numerical simulations consider the discrete nature of the
individuals and progress in discrete time steps. We assume that the
attacker will target the router, among its neighbors, with the
lowest security settings. In addition, we do not allow simultaneous
attacks, so that each infected router will choose its next target
only among those routers which are not already under attack. Once an
attack has started, the attacker will keep trying to bypass the
security setting of the same target until the attempt is finally
successful or not. In both cases, the attacker will then move to
another target. The simulation's unitary time step is defined by the
shortest time scale among all processes involved, i.e. the time
$\tau$ needed to complete an attack to a non-encrypted router with
no password. This automatically defines as unitary the transition
rate $\beta$ associated to the reaction $S_{\nopass}+I\to 2I$, given
$\beta\tau=1$. Typical time scales for the other processes  are:
$\tau_1=$ 6~--~15 minutes to bypass a password in the smaller
dictionary, $\tau_2=$ 400~--~1000 minutes to bypass a password in
the larger dictionary, $\tau_{WEP}=$ 2880~--~5760 minutes to crack
the WEP encryption. The corresponding transition rates can be
analogously defined as probabilities expressed in terms of their
ratio with $\beta$ that defines the unitary rate.

Simulations run for 4,032 time steps, corresponding to 20,160
minutes (i.e., 2 weeks). At each time step we measure the global
attack rate defined as the number of infectious $I(t)$ at time $t$
over the total population of the network discounted by the number of
recovered, $N-R$. In this way, we can take into account for the
differences of the encryption percentages observed in different
urban areas. Given the stochasticity of the process, the numerical
simulations are repeated 100 times changing initial conditions and
assuming different configurations of the randomized network. Average
values and corresponding confidence intervals are also measured.

\bigskip
\noindent {\bf Acknowledgments}

\bigskip
H.H. thanks the Institute for Scientific Interchange in Turin for
its hospitality during the time this work was completed. A.V. is
partially supported by the NSF award IIS-0513650.

\newpage
\lhead{\emph{TABLES AND FIGURES}}

\noindent {\Large{\bf Table and Figure captions}}
\vspace{1cm}

\noindent {\bf Table 1.}    Properties of the giant components of
the WiFi networks for $R_{int}=$45m: size of the giant component
$N$, percentage of encrypted routers $f_{encr}$, maximum degree
$k_{max}$, average degree $\langle k\rangle$, degree fluctuations
$\langle k\rangle^2/\langle k\rangle$. The results presented are
obtained as averages over 5 different randomization procedures to
redefine the location of each router.\\

\noindent {\bf Figure 1.}   {\bf (A)}: Construction of WiFi networks.
Given two routers $i$ and $j$ located at $\vec{p}_i=(LON_i,LAT_i)$
and $\vec{p}_j=(LON_j,LAT_j)$, we place an edge between them if
their distance $d(\vec{p}_i,\vec{p}_j)$ is smaller than the maximum
interaction radius $R_{int}$. {\bf (B)}: Map representation of the
giant components of the WiFi network in the Chicago area as obtained
with different values of $R_{int}$. {\bf (C)}: The degree
distribution for different values of the interaction radius
$R_{int}$ show an exponential decay and a cutoff which depends on
$R_{int}$. The result is obtained as averages over 5 different
randomization procedures to redefine the location of each router.\\

\noindent {\bf Figure 2.} Illustration of the spread of a
wireless worm through Chicago in several time slices. In this
series, the result is based on 1 randomization procedure for the
location of each router and the maximum interaction radius $R_{int}$
is set to 45 meters.\\

\noindent {\bf Figure 3.} {\bf(A)}: Attack rate versus time for the
giant component of all the seven urban areas, keeping $R_{int}=45m$.
{\bf(B)}: The average and 90\% C. I. for three prototypical cases.
{\bf(C)}: The correlation between the final attack rate and average
degree. {\bf(D)}: The correlation between the final attack rate and
degree fluctuations.\\

\noindent {\bf Figure 4.} Compartmental Flows for the Epidemic
Model.\\

\newpage

\captionsetup{labelformat=empty}
\begin{table}
\begin{center}
\vskip .7cm
\caption{\label{table_properties}}
\begin{tabular}{lcccccc}
\hline City & $N$ & $f_{encr}$ & $k_{max}$ & $\langle k\rangle$ &
$\frac{\langle k\rangle^2}{\langle k\rangle}$
&\\
\hline NYC & 36807 &  25.8\%  &  109 &  23.9 &  31.3 \\
\hline Boston & 15899 & 21.7\% &  116 &  21.3  &  35.7 \\
\hline Seattle & 3013 & 26.0\%  & 49  &  11.9 &  16.7\\
\hline Chicago & 50084 & 33.7\% &  154 &  20.3  & 33.6 \\
\hline N.IN & 2629 & 24.3\% &   87 &   18.4 &  29.9 \\
\hline S.IN & 998 &  11.0\% &  32 &  10.4 &  14.1\\
\hline SF Bay & 3106 &  40.1\% &  31  &  9.6  &  12.8\\
\hline\\
\end{tabular}
\vskip .1cm
{\bf Table 1.}
\end{center}
\end{table}

\end{document}